\documentclass[letterpaper,twocolumn,10pt]{article}
\usepackage{usenix2019_v3}

\usepackage{tikz}
\usepackage{amsmath}

\usepackage{comment}
\usepackage{booktabs}
\usepackage{caption} 
\usepackage{xcolor}
\usepackage{color, colortbl}
\definecolor{dark green}{RGB}{94,145,40}
\usepackage{multirow}
\usepackage{xurl}

\usepackage{stfloats}

\newcommand\ChapterPrecis[2]{%
\begin{tikzpicture}[remember picture,overlay]
\node[anchor=north, draw=black, fill=yellow!20, inner sep=5pt, rounded corners, yshift=-#1] at (current page.north) 
{\parbox[t][1.3cm][c]{\textwidth}{\small #2}};
\end{tikzpicture}%
}

\newcommand\mypara[1]{\noindent \textbf{#1}}

\providecommand{\eg}{\emph{e.g.,} }

\begin{document}
%-------------------------------------------------------------------------------

%don't want date printed
\date{}

% make title bold and 14 pt font (Latex default is non-bold, 16 pt)
\title{\Large \bf Token Spammers, Rug Pulls, and Sniper Bots: An Analysis of the\\ Ecosystem of Tokens in Ethereum and in the Binance Smart Chain (BNB)}

%for single author (just remove % characters)
\author{
{\rm Federico Cernera}\\
\text{cernera@di.uniroma1.it}\\
Sapienza University of Rome 
\and
{\rm Massimo La Morgia}\\
\text{lamorgia@di.uniroma1.it}\\
Sapienza University of Rome
\and 
{\rm Alessandro Mei}\\
\text{mei@di.uniroma1.it}\\
Sapienza University of Rome
\and
{\rm Francesco Sassi}\\
\text{sassi@di.uniroma1.it}\\
Sapienza University of Rome
} % end author

%for single author (just remove % characters)
%\author{
%{\rm Your N.\ Here}\\
%Your Institution
%\and
%{\rm Second Name}\\
%Second Institution
% copy the following lines to add more authors
% \and
% {\rm Name}\\
%Name Institution
%} % end author

\maketitle

%%%%% Arxiv Reference
\ChapterPrecis{0.8cm}{If you cite this paper, please use the USENIX Security reference: Federico Cernera, Massimo La Morgia, Alessandro Mei, and Francesco Sassi. 2023. Token Spammers, Rug Pulls, and Sniper Bots: An Analysis of the
Ecosystem of Tokens in Ethereum and in the Binance Smart Chain (BNB). \textit{In 32nd USENIX Security Symposium (USENIX Security 23)}. USENIX Association, Anaheim, CA, 3349–3366. \newline \url{https://www.usenix.org/conference/usenixsecurity23/presentation/cernera}
}
%%%%

%-------------------------------------------------------------------------------
\begin{abstract}
%-------------------------------------------------------------------------------
In this work, we perform a longitudinal analysis of the BNB Smart Chain and Ethereum blockchain from their inception to March 2022. We study the ecosystem of the tokens and liquidity pools, highlighting analogies and differences between the two blockchains. We discover that about 60\% of tokens are active for less than one day. Moreover, we find that 1\% of addresses create an anomalous number of tokens (between 20\% and 25\%).
We discover that these tokens are used as disposable tokens to perform a particular type of rug pull, which we call \textit{1-day rug pull}.
We quantify the presence of this operation on both blockchains discovering its prevalence on the BNB Smart Chain.  We estimate that 1-day rug pulls generated \$240 million in profits.
Finally, we present sniper bots, a new kind of trader bot involved in these activities, and we detect their presence and quantify their activity in the rug pull operations.

\end{abstract}

\section{Introduction}
The cryptocurrency market is loosely regulated~\cite{la2020pump,bartoletti2020dissecting}.
Even if policymakers are moving towards building a safer environment for cryptocurrency investors~\cite{su2020digital}, it is a complex task, and needs time. Meanwhile, blockchain-related technologies evolve fast, and with the birth of the DeFi~\cite{zetzsche2020decentralized} investors begin to move from centralized exchanges (CEX) like Binance to decentralized exchanges (DEX). DEXes are distributed Applications (dApp) for trading that run on-chain powered by smart contracts. While regulating the standard cryptocurrency market is difficult, ruling the on-chain trading platform is even more challenging. Indeed, even if the web interface of a DEX can be shut down~\cite{dexes}, its smart contracts are still reachable and working on the blockchain.

DEX and DeFi dApp were born in the Ethereum blockchain, but DeFi services rapidly pop up on all the blockchains that support smart contracts. Although Ethereum plays a leading role in the DeFi world, with over \$68 billion locked in its smart contracts, the BNB Smart Chain or BSC (former Binance Smart Chain) proposes itself as a faster and cheaper alternative.

Uniswap and PancakeSwap are the two most popular DEXes on Ethereum and BSC. They rely on the Automated Market Maker (AMM) model to handle the trading system. At the basis of the AMM model, there is the concept of liquidity pools, a smart contract that handles two tokens (trading pair) that the user can swap. Unlike CEX, where the platform defines the trading pairs, users can create their pair on DEXes and let the other users use it. However, as we will see in the following, some users abuse this freedom to carry out a series of malicious operations.

In this work, we conduct a longitudinal investigation of tokens and liquidity pools in the Ethereum and BSC blockchains. 
We start by parsing over 3 billion transactions of both blockchains, finding more than 1.3 million tokens and 1 million liquidity pools (Sec.~\ref{sec:the-datasets}).
Then, we reconstruct their lifetime---the time from their creation to their last transfer, discovering that approximately 60\% of the tokens have a lifetime shorter than one day (Sec.~\ref{sec:lifetime}). Hence, we define them as \textit{1-day tokens}. 
A tiny fraction of addresses, just 1\%, is responsible for creating more than 20\% of the tokens (Sec.~\ref{sec:serial}).
%Given their overproduction of tokens, we will call these addresses \textit{token spammers}.
Surprisingly, we also find that the tokens with a very short lifetime are actively traded on liquidity pools.
Albeit this phenomenon is present on both blockchains, it is more widespread on BSC.
Diving into this subset of tokens, we observe that a large fraction of liquidity pools used to trade the 1-day tokens show a malicious pattern that we call \textit{1-day rug pull} (Sec.~\ref{sec:case_study}). 
We analyze all the liquidity pools looking for this pattern, and we find 272,349 potential rug pulls on BSC and 21,742 on Ethereum. We estimate the cost of the operation and the gains of the organizers, finding that they earned approximately \$240 million with such activity (Sec.~\ref{sec:gains}). 
Here, we see that the success rate of the 1-day rug pull is not very high (between 40\% and 60\%). However, given the simplicity and the very low cost of the operation, attackers can serially arrange the rug pulls and cover a series of unsuccessful operations with a single successful one. Finally, we study how this kind of operation evolved over time, discovering that the BSC has gradually surpassed Ethereum in terms of the number of operations and gains. Moreover, we find that the operations are more prevalent during two specific events: the 2020 Defi Summer and the 2021 Altcoin season (Sec.~\ref{sec:longitudinal}).

Our key contributions are:
\begin{itemize}
    \item \textbf{Analysis of BNB smart chain}: To the best of our knowledge, we are the first to study this young but well-established blockchain, performing a longitudinal analysis from its inception to March 2022. We study the tokens and the liquidity pools ecosystem, highlighting analogies and differences with Ethereum.
    \item \textbf{Short lifetime tokens and Token spammers}: We estimate the lifetime of the tokens on both blockchains. Discovering that about 60\% of tokens last less than one day.
    Analyzing who creates the tokens, we observe that just 1\% of addresses create an abnormal number of tokens (about 20-25\% of tokens of the blockchains).
    \item \textbf{1-day rug pulls}: We investigate the presence of the rug pull pattern in 1-day tokens. We discover that on BSC, 81.2\% of 1-day tokens listed on PancakeSwap have this pattern.
    We estimate the gains of the attackers, observing that even if the operation is very simple to arrange, given its cheap cost, it is profitable when performed serially.
    \item \textbf{The sniper bot 2.0}: 
    We find the presence of sniper bots (Sec.~\ref{sec:sniper_bots}), a particular kind of trader bot that observes the blockchain's mempool to buy newly listed tokens. To the best of our knowledge, we are the first to illustrate how this kind of trading bot works, detect their presence, and quantify their activity in the rug pull operations.
\end{itemize}
\section{Ethereum and BNB Smart Chain}
\label{sec:background}

Ethereum~\cite{buterin2014next} is a 
proof-of-work\footnote{Switched to proof-of-stake from 2022-09-15}
blockchain. Its native coin is the Ether (ETH), the second most popular cryptocurrency after Bitcoin (BTC), with a market cap of more than 210 billion USD.
A key feature of Ethereum is smart contracts, pieces of code that execute in a decentralized way on-chain.
%Smart contracts enable the creation of decentralized applications (dApp) and the so-called Web3.0~\cite{alabdulwahhab2018web}.
Through smart contracts, it is possible to create new digital assets like (fungible) tokens and NFTs (non-fungible tokens).
\\
\textbf{The tokens.}    
Tokens are cryptocurrencies that can be exchanged or traded. They are created on top of the blockchain, and their mechanisms are defined using smart contracts.
On Ethereum, the ERC-20~\cite{erc-20} standard defines the main properties of tokens. An ERC-20 compliant smart contract must implement a set of functions and events specified in the standard. These functions are reported in Table~\ref{tab:erc20signature}. Some of them are optional, in particular the \textit{name()}, the \textit{symbol()}, and the \textit{decimal()} functions.
On Ethereum, tokens and digital assets are held into accounts.
\\
\textbf{Ethereum accounts.}
There are two kinds of accounts on Ethereum: Externally owned accounts (EOA) and contract accounts.
EOAs consist of a pair of public and private keys generated with the  Elliptic Curve Digital Signature Algorithm (ECDSA)~\cite{johnson2001elliptic}. An account is represented by its public address, a 42-character hexadecimal string obtained concatenating "0x" to the last 20 bytes of the Keccak-256~\cite{dworkin2015sha} hash of the public key.
Instead, a contract account is tied to a smart contract. It is represented with an address in the same format as an EOA. A contract account is generated when a smart contract is deployed to the Ethereum blockchain.
\\
\textbf{Transactions and fee.}
A transaction is an action that updates the whole Ethereum network.
It can be used to move digital assets, deploy a smart contract, or invoke a smart contract.
Executing a transaction has a cost, commonly called transaction fee. The fee is variable and depends on two main factors: The state of the network (if the network is heavily loaded, the fee is usually higher) and the complexity of the operation that the transaction triggers.
\\
\textbf{Smart contract deployment.} 
Smart contracts are programs that run on the Ethereum blockchain. They are written in a high-level programming language (\eg Solidity~\cite{dannen2017introducing}) and compiled into bytecode that runs on the Ethereum Virtual Machine (EVM)~\cite{evm_eth}.
A smart contract can be deployed by sending a contract creation transaction from an EOA to the zero address\footnote{0x0000000000000000000000000000000000000000}.
The contract creation transaction contains the bytecode of the smart contract.
A smart contract can also create new smart contracts.
Since a smart contract can start a transaction only in response to a transaction that triggers it, an EOA must trigger the generation of a new smart contract.
\\
\textbf{Events and logs.}
To facilitate the tracking of internal states of smart contracts, Ethereum provides Events and an internal Logs register.
Each time an action changes the internal state of a smart contract, it can fire an Event that notifies the change. 
%All the events are written on an Event log. \mlnote{Thanks to it, users and developers can easily track the state of the smart contracts in the blockchain. --- potrebbe essere eliminato} AM Eliminato
\\
\textbf {EVM and EVM compliance.} 
Ethereum is a distributed state machine that changes its state at each new block accordingly to a predefined set of rules. 
The EVM~\cite{wood2018ethereum} is the entity that computes these changes in states.
Other than Ethereum, other blockchains rely on the EVM (\eg BNB Smart Chain~\cite{bsc}, Avalanche~\cite{avalanche}, Fantom~\cite{fantom}), and they use one of the standard EVM or a custom one. 
These blockchains are called EVM-compliant. They run the same (or with minimal change) smart contract written for Ethereum, use the same convention for the address, and handle states the same way as Ethereum.
\\
\mypara{The BNB Smart Chain.}
The BNB Smart Chain~\cite{bsc} (previously Binance Smart Chain) or BSC is a blockchain that was born in 2020.
%as a parallel to the Beacon Chain (previously Binance Chain), and together they form the BNB Chain.
Its consensus is based on the PoSA~\cite{poa} (Proof of Stake and Authority).
The coin of both chains is the BNB (Build and Build, previously Binance Coin)---the third coin by market cap with over 46 billion of capitalization.
As Ether for Ethereum, the BNB coin fuels the transactions on the BNB chain. 
Because of EVM compatibility, it is possible to create tokens in BSC similarly to Ethereum. However, in this case, tokens follow the BEP-20 standard instead of the ERC-20.

\section{AMMs, Uniswap and its forks}
Uniswap~\cite{adams2020uniswap} is one of the most popular decentralized applications (dApp).
According to DefiLlama~\cite{defilama}, a popular DeFi statistics aggregator, Uniswap is the $5^{th}$ dApp by TVL (Total Value Locked, amount of money locked into smart contracts) with over 6 billion USD.

Uniswap is the first dApp to use the \textit{AMM model} successfully. This model relies on a mathematical formula to fix the price of assets and on the concept of liquidity pools and providers. A \textit{liquidity pool} is a smart contract that contains two or more cryptocurrencies that the user can swap for another. Instead, a liquidity provider is a user who invests in the liquidity pool, providing cryptocurrencies to the smart contract. 
When a liquidity provider injects liquidity into the liquidity pool, the smart contract mints LP-tokens and gives them to the liquidity provider.
The LP-token represents the share of the liquidity pool owned by the investor. When the liquidity provider desires to get back his cryptocurrencies, he transfers the LP-tokens to the smart contract.
The latter burns the LP-tokens and provides the cryptocurrencies back to the investor.
In Uniswap V2, each pool consists of a pair of ERC-20 tokens.
The liquidity pool is divided into two parts, each containing a single token, and both have an equivalent value. Let a pool consist of $x$ token $A$ and $y$ token $B$. At each swap, the pool preserves $x*y$. When a user swaps $a$ token $A$ for token $B$ (the user adds token $A$ to the pool and takes token $B$ from the pool), $x$ increases by $a$ and $y$ decreases by $b$, where $b$ is computed so that $x*y$ does not change. The rate $a/b$ of the exchange depends on the ratio of $x$ and $y$ in the pool. Consequently, the swap operation changes the current exchange rate. The value of token $A$ decreases while the value of token $B$ increases, and the two parts maintain the same value.

Because of the success of this model, the popularity of Uniswap and 
its open-source smart contracts~\cite{uniswapv2license}, more than 50 protocols were born on several blockchains by forking Uniswap smart contracts in the last years.
Uniswap is on its third version, but all its forks belong to the second version since the third one is under a Business Source License. For this reason, in this work, we focus on Uniswap V2 and its forks.
%https://uniswap.org/blog/uniswap-v3
One of the most popular forks of Uniswap is PancakeSwap, which lives on BSC. It is the $1^{th}$ dApp by TVL on this blockchain with over 4 billion USD locked in its smart contracts.

\section{The Datasets}
\label{sec:the-datasets}

For our investigation, we build two different datasets: 
The \textit{Token Dataset}, which contains all the ERC-20 (resp. BEP-20) tokens created, and the \textit{Liquidity Pool Dataset}, which contains data about liquidity pools. Each dataset has two versions, one with data from the Ethereum blockchain and the other from the BNB Smart Chain.

We consider the whole history of both blockchains from their inception to March 2022. For the Ethereum blockchain, we process all the blocks from block 0 (2015-07-30) to block 14340000 (2022-03-07). For the BSC blockchain from block 0 (2020-04-20) to block 15854000 (2022-03-07).
Given the large amount of data and the need to parse the entire blockchains multiple times, for performance reasons and to avoid overloading public nodes (\eg nodes provided by Binance~\cite{bscrpc} and Infura~\cite{infura}) or services (\eg BscScan or Etherscan), we host and run an Ethereum and a BNB Smart Chain node. Finally, to query the blockchains and process the data, we use the Web3~\cite{web3} and the Ethereum-etl~\cite{ethereumetl} Python libraries.
Web3 is a collection of libraries that allow the interaction with a local or remote EVM-compliant node. Ethereum-etl allows extracting information from EVM-compliant blockchains and exporting it into formats like CSV or JSON.
The data collection phase was performed on an Ubuntu 20.04 machine with AMD EPYC 7301 (16-Core Processor, 2.80 GHz), 1 TB of RAM, and  4 TB SATA SSD with 560/530 MB/s read and write speed. Data processing took between 24 and 72 hours each time we parsed the entire blockchain, depending on the kind of data retrieved.

%\fcnote{By using the amount of transactions given by EtherScan~\cite{etherscan-api} and BscScan~\cite{bscscan-api} as our ground truth, we were able to verify that we had successfully recovered all the transactions that occurred during these times on the two blockchains.}

\subsection{The Token dataset}
\label{sec:token-dataset}

\subsubsection{Gathering smart contracts}
\label{sec:gathering}
As a first step to building the Token dataset, we collect all the contract creation transactions issued by EOAs. As mentioned in Sec.~\ref{sec:background}, EOAs can deploy a smart contract by sending a \textit{contract creation transaction} to the zero address. We process all the transactions in the considered time frame in BNB Smart Chain (2.6 billion transactions) and in Ethereum (1.4 billion transactions). We collect 2,195,399 and 4,420,389 contract creation transactions respectively.

However, tokens can also be created by a smart contract itself. Indeed, it could be the case that an EOA calls a smart contract method, and its execution generates a new ERC-20 (or BEP-20) compliant smart contract. In this case, the token is created with a so-called \textit{internal transaction}. Despite the name, internal transactions are not real transactions, but rather calls performed by smart contracts. These kinds of transactions are stored off-chain---they are not visible simply parsing the blockchain.

To track the tokens created by internal transactions, we can operate in two ways: The first way is to re-execute all the transactions in the blockchain in the EVM and trace all the calls. This process is extremely expensive~\cite{evm_expensive} from a computational point of view. The alternative is to scan the Event log looking for events that emit a \textit{Transfer event}. The second way is much faster and we estimate that it loses only 12\% of the total number of tokens created by internal transactions. Moreover, the missing tokens are never been used, traded or transferred, and are thus of little importance for our study (we discuss in detail the impact of this choice in Sec.~\ref{sec:discussion}). So, we parse all the logs of both blockchains, searching for smart contracts that emit a Transfer event compliant with the ERC-20 (resp. BEP-20) interface. Then, we use Etherscan~\cite{etherscan-api} and BscScan~\cite{bscscan-api} to retrieve the transactions that created these smart contracts and all the information.

At the end of these two steps, we have a collection of 3,087,274 and 4,534,599 smart contracts extracted from BSC and Ethereum, respectively. For each of them, we store the following information: The \textit{address of the contract}, the \textit{block number} in which the smart contract has been generated, the block in which the smart contract emits its last event, the EOA that deployed the smart contract or in the case of internal transactions the EOA address that triggers the first smart contract, the amount of \textit{gas used}, the cost of the gas unit (\textit{gas price}), the \textit{bytecode} of the smart contract, and if the smart contract has been deployed by an EOA or through an internal transaction. 

\subsubsection{Token identification}
Smart contracts are not only used to create tokens, as well as not all smart contracts that emit a Transfer event are tokens (\eg NFT contracts).
Thus, we need to identify which of the retrieved smart contracts are ERC-20 (resp. BEP-20) compliant. Unfortunately, this is not a trivial task, and in the last years several works~\cite{di2021identification, victor2019measuring, chen2020traveling,chen2019tokenscope,frowis2019detecting}, attempted to face this problem with several approaches that we describe in Sec.~\ref{sec:sota}.
For our analysis, we follow the approach proposed by ~\cite{victor2019measuring,chen2020traveling} that leverage the bytecode of smart contracts.

According to the Solidity specification~\cite{ABIspec}, in the bytecode, smart contract's methods are identified by signatures that consist of the first 4 bytes of the \textit{Kekkack-256} hash of the method name and parameters' type.
Thus, to verify if a bytecode of a retrieved smart contract represents an ERC-20 (resp. BEP-20) compliant token, we verify if it contains at least all the signatures of the ERC-20 (resp. BEP-20) mandatory methods.
Tab.~\ref{tab:erc20signature} in the Appendix shows the signature of the mandatory and optional methods of the ERC-20 and BEP-20 interfaces.

Of the 4,534,599 smart contracts' bytecodes retrieved on the Ethereum blockchain, we find that 389,348 (8.5\%) are ERC-20 tokens compliant, and 381,551 (98\%) of them also implement the optional functions of the ERC-20 interface.
Instead, on the BNB Smart Chain, we find that 1,887,484 out of 3,087,274 (61\%) are BEP-20 compliant, and, as for Ethereum, almost all of them also implement the optional methods of the BEP-20 interface. Although we found more smart contracts on Ethereum than in BSC (4,534,599 vs. 3,087,274), there are many more compliant tokens in BSC (1,887,484) than in Ethereum (389,348). This discrepancy suggests that BSC may be a more interesting environment to study tokens and, possibly, their misuse.

Lastly, we retrieve all the information about the identified tokens such as the name, the symbol, the number of decimals, and the total supply. To do so, we use the Ethereum-etl library and the Contract Application Binary Interface (ABI)~\cite{ethereum-abi}. The ABI is an interface between two program modules. It contains the specification for encoding/decoding methods and structures to interact with the machine code and interpret the results. Through the library, it is possible to instantiate smart contracts in an object-oriented manner and call its methods using an appropriate ABI. We instantiate the token contracts using an ABI that contains the specifications of ERC-20 (resp. BEP-20) methods and call the  \textit{name()}, \textit{symbol()}, \textit{decimals()}, \textit{totalSupply()} methods.

At the end of the process,  we have a dataset of ERC-20 (resp. BEP-20) tokens containing all the information about the smart contracts described in Sec.~\ref{sec:gathering} and the related tokens. Table~\ref{tab:erc20} shows the number of smart contracts on both blockchains.

\begin{table}
\small
    \centering
    \caption{%
       An overview of the Token dataset.
    }\label{tab:erc20}
    \begin{tabular}{@{\extracolsep{2pt}}l r r r r}
        \toprule
        {} & \multicolumn{2}{c}{Ethereum}  & \multicolumn{2}{c}{BNB Smart Chain}\\
        \cmidrule{2-3} 
        \cmidrule{4-5} 
        Contracts & Total  & ERC-20 & Total  &  BEP-20 \\
        \midrule
        External   & 4,420,389 & 293,688 & 2,195,399 & 1,021,427 \\ 
        
        Internal & 114,210 & 95,660 & 891,875 & 866,057 \\
        \midrule
        Total &  4,534,599 & 389,348 & 3,087,274 & 1,887,484 \\
        Total (w/o LP) &  - & \textbf{323,863} & - & \textbf{1,078,016}
 
    \end{tabular}
\end{table}

\subsection{Liquidity Pools dataset}
\label{sec:liquidity-dataset}
 \begin{table}
\small
    \centering
    \caption{%
       An overview of the Liquidity pools dataset.
    }\label{tab:liquidity_pool_dataset}
    \begin{tabular}{@{\extracolsep{2pt}}l r r r r}
        \toprule
        {} & \multicolumn{2}{c}{Ethereum}  & \multicolumn{2}{c}{BNB Smart Chain}\\
        \cmidrule{2-3} 
        \cmidrule{4-5} 
        Events & Uniswap  & Others & PancakeSwap  & Others \\
        \midrule
        PairC.   & 65,098 & 5,483 & 941,220 & 30,907 \\ 
        Mint & 1,399,599 & 512,319 & 21,944,474 & 5,027,980 \\
        Burn & 824,359 & 243,482 &  7,339,286 & 2,481,023 \\
        Swap & 54M & 27M & 571M & 179M \\

    \end{tabular}
\end{table}

To create the Liquidity Pool dataset, we consider Uniswap, its forks, and the other protocols that leverage its smart contracts.
%In the following, we explain the main smart contracts of Uniswap that also hold for its forks.

Uniswap has three main smart contracts: \textit{Factory}, \textit{Pair}, and the \textit{Router}.
The Factory contract is responsible for creating the smart contract that handles the liquidity pool and the LP-tokens. %\mlnote{Note that since the LP-token and the liquidity pool are handled by the same smart contract they have the same contract address.}
The Pair contract keeps track of the balances of the tokens in the pool and implements the AMM logic explained in Sec.~\ref{sec:background}. The Router contract offers the entry point to interact with the liquidity pools.  Thus, it is possible to swap tokens and add or remove cryptocurrencies from a liquidity pool by interacting with the Router. 
Each of these contracts implements a set of Events that notify their status changes. 

To build our datasets, we parse the Event log of the Ethereum and BSC blockchains.  Following, we report the events we look for and a brief description:
\begin{itemize}
    \item \textbf{PairCreated:} This event is fired by the Factory contract each time a new liquidity pool is created. We find 972,127 and 70,581 PairCreated events emitted on BSC and Ethereum, respectively. From the event, we can obtain the transaction hash, the block of the creation of the liquidity pool, the address that created the liquidity pool,  the address of the liquidity pool, and the addresses of the two tokens (the pair of the liquidity pool), the gas used and the price paid per gas.
    Analyzing the address that fired the event, we find that almost all the liquidity pools of BSC are created in PancakeSwap (96.8\%), and almost all the liquidity pools of Ethereum are created in Uniswap (92.2\%).
    %\mlnote{Analyzing the address that fired the event and looking online for notable smart contract addresses, it is possible to have a rough idea of the diffusion of the Uniswap forks in the blockchains.
    %In BSC, we find that PancakeSwap created most liquidity pools, with 941,220 emitted events (96.8\%), followed by ApeSwap~\cite{apeswap} (3,265 events), BakerySwap~\cite{bakeryswap} (2,418 events) and  Mdex~\cite{mdex} (1,602 events).
    %In Ethereum, Uniswap emitted 65,098 events (92.2\%), while the  SushiSwap~\cite{buterin2014next} Factory contract, a popular alternative to Uniswap on Ethereum, 2,637 (3\%).---secondo me interessante ma probabilmente inutile ai fini del paper }

    \item \textbf{Mint \& Burn}:  The Pair contract emits a Mint (or Burn) Event each time an LP-token is minted (or burned). This occurs whenever a liquidity provider adds (or removes) tokens into a liquidity pool.
    Analyzing these events, we obtain the transaction hash and the block of the Mint (Burn) Event, the address of the liquidity pool, the address that added (removed) the liquidity, the number of LP-tokens minted (burned), the gas used, and the price paid for the gas.
    We find 26,972,454 Mint events and 9,820,309 Burn events on BSC, and 1,911,918 Mint events and 1,067,841 Burn events on Ethereum.
   
    \item \textbf{Swap:} This event is fired by the Pair contract each time a user swaps tokens in a liquidity pool. From the event, we obtain all the information related to the swap: The transaction hash, the block in which the swap occurs, the address of the liquidity pool used, the address that performs the swap, the number of tokens swapped, the gas used and the gas price. We find 750,508,160 events on BSC and 82,447,051 events on Ethereum.
    
\end{itemize}
Moreover, we complete our dataset collecting for each smart contract the block number in which it emits the last event.
Tab.~\ref{tab:liquidity_pool_dataset} describes the final dataset. 

Given that LP-tokens are ERC-20 (resp. BEP-20) compliant tokens, they are already present in our Tokens Dataset. However, our goal is to study standard tokens and liquidity pools separately. Thus, as the final step, we get rid of the information related to the LP-tokens from the Tokens Dataset.
The last line on Tab.~\ref{tab:erc20} reports the number of tokens after we get rid of the LP-tokens.

\section{The Lifetime of tokens}
\label{sec:lifetime}
Our data collection revealed a surprisingly high number of tokens and liquidity pools on Ethereum and BSC. Services like CoinGecko~\cite{coingeckoapi} or CoinmarketCap~\cite{coinmarketcap} list about 13,000 cryptocurrencies on 602 centralized and decentralized exchanges. Therefore, it is unclear what is the role of the large majority of tokens in the blockchain ecosystem.

%To obtain this information, we leverage CoinGecko~\cite{coingeckoapi}, a service that records trading data for more than 13 thousand cryptocurrencies and tokens on 602 centralized and decentralized exchanges.
%We use CoinGecko APIs to retrieve all the tokens created on the BNB Smart Chain and Ethereum. 
%We found 4,534 tokens created exclusively on BSC, 4,381 tokens created exclusively on ETH and 791 tokens created on both platform.
%\subsection{Lifetime analysis}
%\label{sec:lifetime}

\begin{figure}
    \centerline{\includegraphics[width=0.50\textwidth]{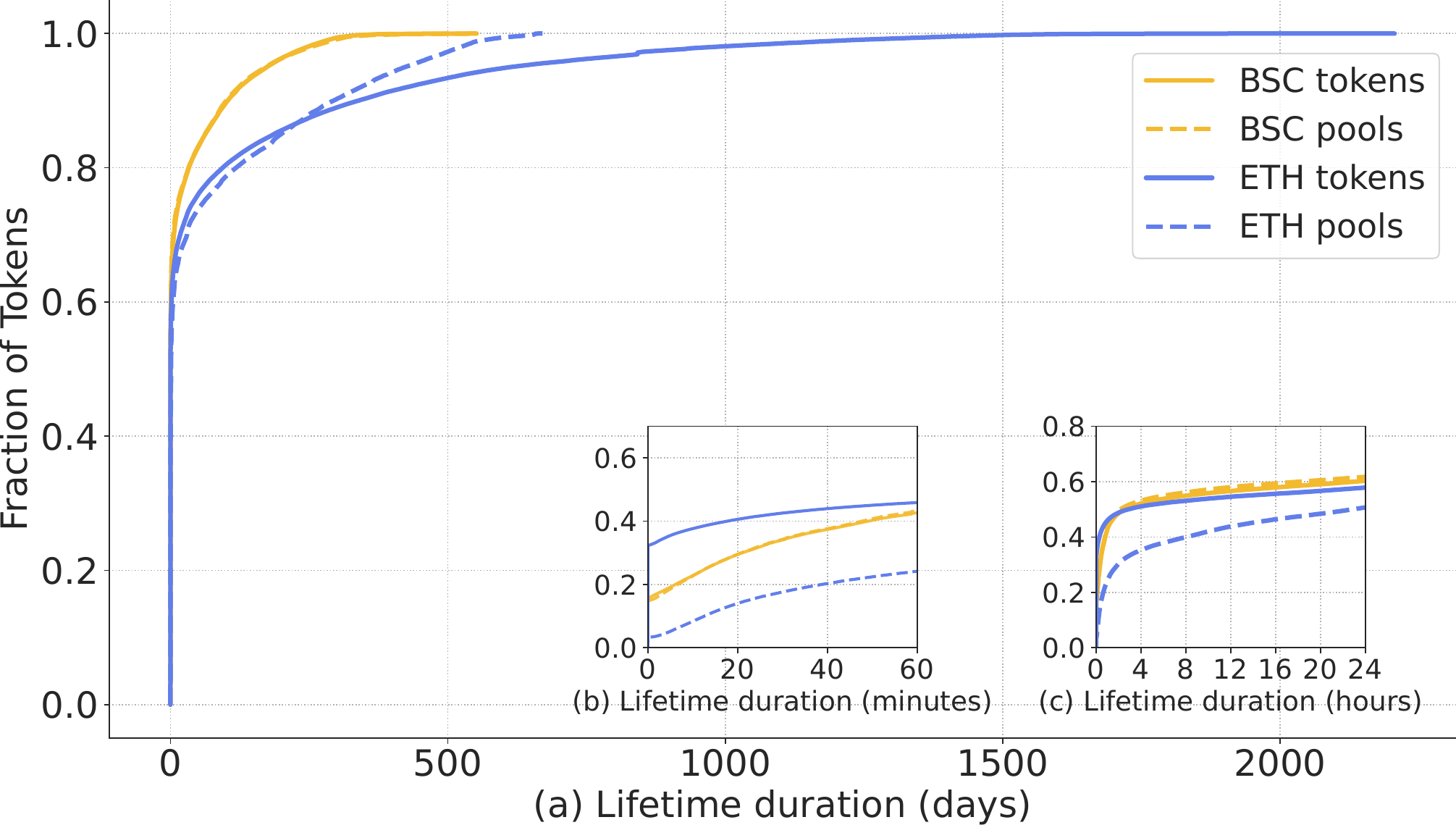}}
\caption{Lifetime of tokens and liquidity pools on BSC and Ethereum.}
  \label{fig:cdf_lifetime}
\end{figure}

To obtain a first insight into the characteristics of tokens and liquidity pools, we introduce the concept of \textit{lifetime}.
We define the lifetime of a token in the following way: A token begins its lifetime at the block where its smart contract has been deployed, while it ends its lifetime in the last block where it emits any Event.
Similarly, a liquidity pool begins its lifetime at the block where the PairCreated event is emitted, and it ends in the last block where the liquidity pool' smart contract emits any Event.

Fig.~\ref{fig:cdf_lifetime} shows the CDF of tokens' lifetime and liquidity pools' lifetime on Ethereum (blue lines) and BSC (yellow lines). Tokens and liquidity pools are shown with solid and dashed lines, respectively. The slope of the curves tells that the lifetime of the tokens in BSC is generally shorter than the lifetime of the tokens in Ethereum.
Consider that BSC is a young blockchain, with slightly more than two years of activity (released on 2020-04-20), while Ethereum is more than seven years old (released on 2015-07-30). The longevity of Ethereum is also visible by the long tail of its tokens in the CDF. Nonetheless, it seems that Ethereum's tokens that tend to be more solid and long-lasting. This difference is smaller when we look at liquidity pools. Indeed, PancakeSwap, which handles about 97\% of the liquidity pools in BSC, was born only four months after the release of Uniswap V2.
From the CDF, we can also note a few additional interesting facts, particularly when we look at the first 24 hours of the life of tokens and liquidity pools.
%Indeed, the BSC was released in ap recently (20-04-2020) than Ethereum (30-07-2015).
%A fraction of Ethereum tokens have a lifetime that lasts for years, while BSC tokens have a maximum lifetime of about one and a half year. This is expected since the BSC was released more recently (20-04-2020) than Ethereum (30-07-2015).
%Lifetimes of liquidity pools are generally shorter than the lifetimes of tokens on both blockchains.
%This is not surprising. For a liquidity pool to exists for a token, the token must have been created in advance. Thus the lifetime of the token have to be longer than the lifetime of a liquidity pool that contains that token.\fsnote{forse bisogna specificare che quando un token è in una liquidity pool noi lo vediamo comunque attivo perché emette eventi.}

%\mypara{BSC tokens are created to be used in liquidity pools.} Looking at the gap between the solid and dashed lines we can see that the gap is much smaller for BSC than for ETH. This means that the life of liquidity pool in BSC is almost the same as the duration of the lifetimes of tokens, while for Ethereum tokens last on average longer than the liquidity pools. This suggests that tokens in BSC are created only to be used in liquidity pools and thus they are created very closely to the creation of the liquidity pool and cease to be active when the liquidity pool is "closed". To confirm this intuition we study the distance between the creation of a token and its appearance in a liquidity pool.

\mypara{A significant fraction of tokens is never active.} Looking at the zoomed image in the center of Fig.~\ref{fig:cdf_lifetime}~(b), it is possible to see that a significant fraction of tokens have a lifetime of length zero, meaning that the token is active only in one block, when it was created. This phenomenon is more common in Ethereum, with 104,836 out of 323,863 (32.4\%) tokens that belong to this category, against 167,318 out of 1,078,016 (15.5\%) in BSC.
In the following, we refer to the tokens that last only one block as \textit{1-block tokens}, while to the other tokens as \textit{active tokens}. We find 910,698 and 219,027 active tokens on BSC and Ethereum, respectively. Table~\ref{tab:bsc_eth_short_lifetime} succinctly reports on these statistics. 

\mypara{A large part of active tokens has an extremely short lifetime.}
Fig.~\ref{fig:cdf_lifetime}~(c) shows that about 60\% of the tokens in BSC and Ethereum have a lifetime shorter than one day. We refer to these tokens as \textit{1-day tokens}. Considering only active tokens, we find that 471,385 (51.7\%) of all the active BSC tokens and 82,542 (37.7\%) of all the Ethereum active tokens are 1-day tokens. Looking at the data at a higher granularity (Fig.~\ref{fig:cdf_lifetime}~(b)), we can note that the death ratio of BSC tokens is surprisingly high.
Proportionally, BSC has approximately half of the 1-block tokens of Ethereum, about the same proportion of dead tokens after 60 minutes, and a significantly larger proportion of dead tokens after the first 4 hours.
As we can see in Fig.~\ref{fig:cdf_lifetime}~(c), the first four hours of token life are also crucial in Ethereum.

\mypara{Almost all the BSC tokens with short lifetimes have a liquidity pool.} 
Here, we find one of the main differences between BSC and Ethereum.
468,556 out of 471,385 (94.8\%) active tokens with a lifetime shorter than one day in BSC have a liquidity pool. In Ethereum, only 33.1\% (27,346). It seems that on BSC the liquidity pool is the main reason for creating a token.

\begin{table}
   \centering
   \caption{%
       Summary of 1-day and 1-block tokens for BSC and Ethereum. 
  }\label{tab:bsc_eth_short_lifetime}
   \begin{tabular}{l l  r r r}
      \toprule
      Lifetime & BSC & Ethereum \\
      \midrule
      1-day & 638,703 (59.2\%) &  187,378 (57.8\%) \\
      %One-hour & 452,708 (42.0\%) & 115,616 (47\%)\\
      1-block & 167,318 (15.5\%) & 104,836 (32.4\%)\\
      \midrule
      Total tokens & 1,078,016 &  323,863\\
      
       \end{tabular}
\end{table}

%\begin{table}
%   
%   \centering
%   \caption{%
%       Summary of 1-day and 1-block tokens created by token spammers on BSC and %Ethereum. 
%  } \label{tab:bsc_eth_token_spammers}
%   \begin{tabular}{l r r r r}
%      \toprule

%      Lifetime & BSC & Ethereum \\
%      \midrule
%      1-day & 170,768 (69.5\%) & 40,552 (59.7\%)\\
%      %One-hour & 147,268 (45.5\%) & 34,438 (50.7\%)\\
%      1-block & 29,829 (12.1\%) & 25,185 (37.1\%)\\
%      \midrule
%      Total tokens (spammers) & 245,554 &  67,869\\
%       \end{tabular}
%\end{table}

\section{Token spammers}
\label{sec:serial}
\begin{figure}
    \centerline{\includegraphics[width=0.50\textwidth]{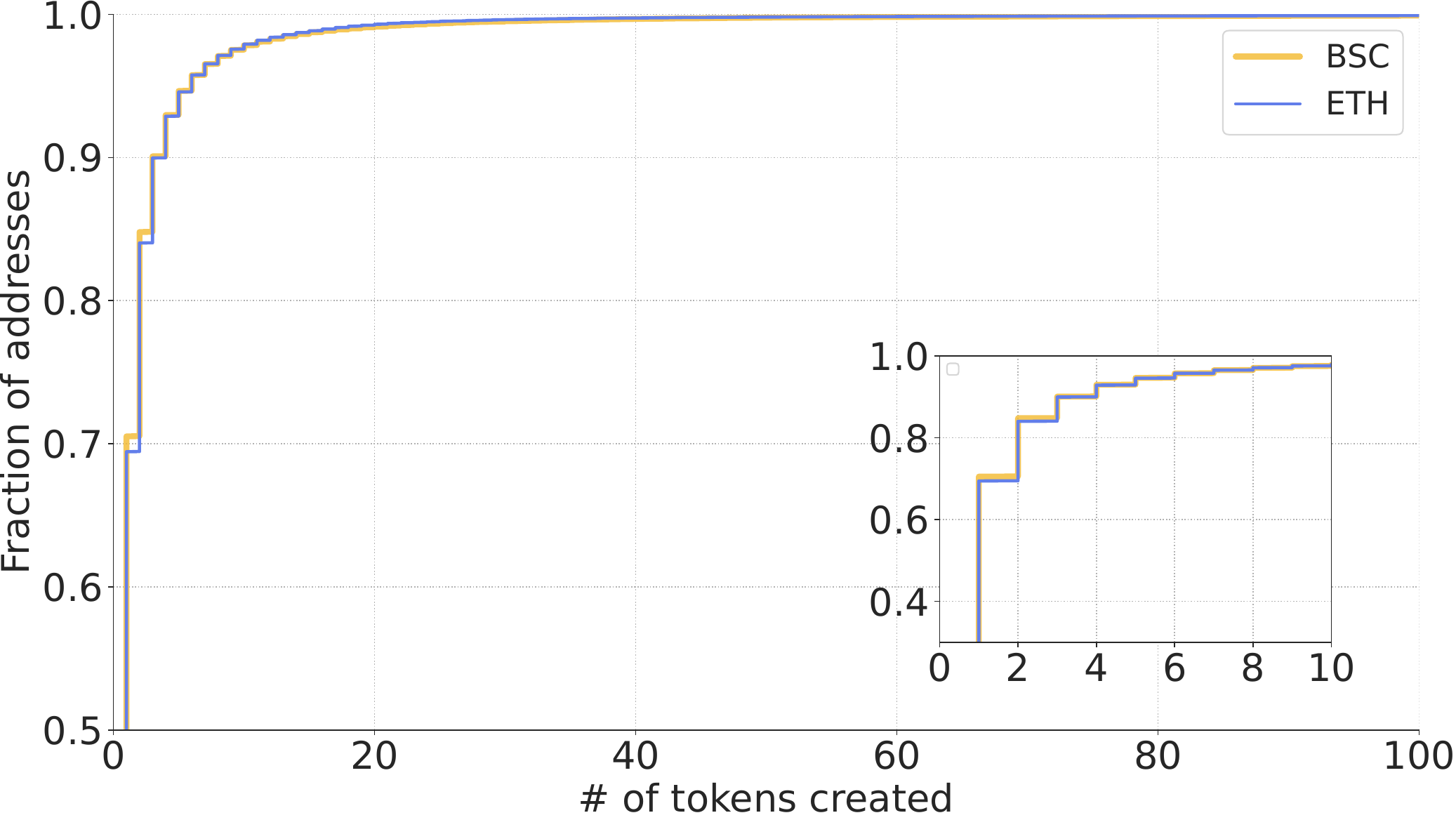}}
\caption{Distribution of the number of tokens created by the addresses that create at least one token in BSC and Ethereum. For the sake of visualization, the CDF is cut at 100 tokens.}
  \label{fig:addresses_distribution}
\end{figure}

In this section, we change perspective and explore who creates tokens.
Retrieving the list of creator addresses from our token dataset, 
we find 144,795 and 464,095 different addresses that create at least one token, respectively, in Ethereum and BSC. Comparing these numbers with the total number of cumulative unique addresses on Ethereum (189,858,744) and BSC (140,522,222)\footnote{Data retrieved from Etherscan and BSCscan respectively}, we see that they represent only a very small fraction of the addresses, the 0.07\% in Ethereum and 0.33\% in BSC.
Fig.~\ref{fig:addresses_distribution} shows the distribution of the number of tokens created by addresses in Ethereum and BSC. The first thing to notice is that the two distributions are extremely similar. The large majority of these addresses (70\%) create only one token, as we can see in the zoomed image on the bottom right corner of Fig.~\ref{fig:addresses_distribution}. 95\% of addresses create five tokens or less, and just 1\% of addresses create more than 18 tokens.
%However, we can gather further insights by plotting the same data differently.

\begin{figure}
    \centerline{\includegraphics[width=0.50\textwidth]{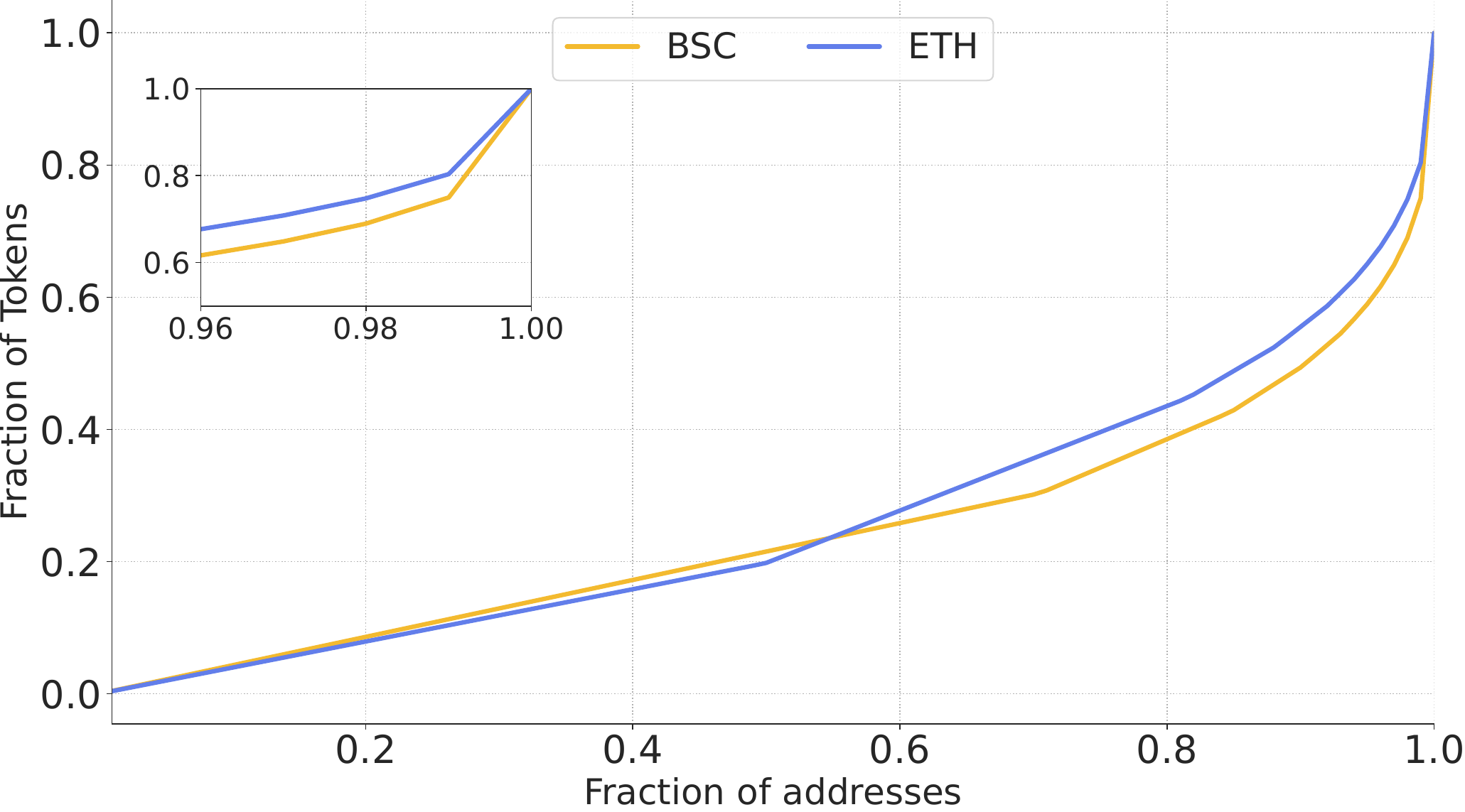}} \caption{Fraction of addresses that create at least one token with respect to the fraction of tokens that they create.}
  \label{fig:token_distribution}
\end{figure}

\mypara{A small fraction of addresses creates a disproportionate amount of tokens.} Fig.~\ref{fig:token_distribution} shows the CDF of tokens created by fraction of addresses.
From the figure, we can see that although 70\% of addresses create just one token, the total amount of tokens created by these addresses account for only 30\% of the tokens on the two blockchains.
And more interestingly, we find that just 1\% of the addresses
creates 24.3\% (262,023) of the tokens in BSC, and similarly, 1\% of the addresses in Ethereum create 20.1\% (67,869) of the tokens.
These addresses create an average of 51 and 61 tokens in Ethereum and BSC, respectively. We will refer to these addresses as \textit{token spammers}. 

%\subsection{The token spammers}
%\label{sec:token_spammers}
%In this subsection we put under the lens the token spammers. Here are a few interesting observations.

\mypara{Token spammers are more prevalent in BSC.}
Although the distribution of the number of tokens created by addresses in Ethereum and BSC is almost identical (Fig.~\ref{fig:token_distribution}), the absolute numbers are different. Indeed, in terms of raw numbers, we find that BSC has almost four times more token spammers than Ethereum (4,231 vs. 1,329), and the spammers of BSC create almost four times more tokens in BSC than in Ethereum (262,023 vs. 67,838). %This is not surprising since we find more tokens on BSC than in Ethereum (387,360 vs 1,864,979).

\mypara{Token spammers create tokens mainly with contract creation transactions.} As mentioned in Section~\ref{sec:background}, tokens can be created in two ways: By sending a contract creation transaction or by sending a transaction to a smart contract that generates the token.
We find that 94.8\% of the tokens on BSC and 82.3\% of the tokens on Ethereum are created directly by sending a contract creation transaction.

\mypara{Token spammers create short lifetime tokens.} 
As we have seen, a significant amount of tokens have a lifetime shorter than one day. Investigating the relationship between token spammers and 1-day tokens, we discover that most of the tokens created by the spammers have a lifetime shorter than one day. The spammers created 170,768 1-day tokens out of 262,023 (65.1\%) and 40,552 1-day tokens out of 67,869 (59.8\%), respectively, in BSC and Ethereum.

\section{The Anatomy of a Rug Pull}
\label{sec:case_study}
The top token spammer creates 17,936 tokens in the timeframe of our analysis. If we look at the name of these tokens, we find that almost all of them have the same name (the tokens have only 76 unique names), with the most used being 'Pornhub' with 605 occurrences. The median lifetime of these tokens is extremely small: 45 mins. Lastly, almost all of the tokens (99.7\%) created by this address have a liquidity pool. 
We study the liquidity pools of these tokens and find out that they are used to perform an operation commonly known as rug pull~\cite{rugpulls,mazorra2022not}. In the following, we report a detailed example of a rug pull operation carried out by this address.

We focus on \textit{OnlyFans}\footnote{0xe8b6f08841d668605343A63144D76ff2dE9A1199}, a token created by the top token spammer on block 8090747 
(2021-06-07 01:40:34 PM UTC) by issuing a contract creation transaction. %This token has a supply of $10^{13}$ and its symbol is the unicode \textit{U+128139}, the emoji of a kiss mark, followed by the string "OnlyFans". 
On block 8090751 (2021-06-07 01:40:46 PM UTC), after 4 blocks from its creation, the token spammer creates a liquidity pool that contains the pair (OnlyFans, WrappedBNB) and adds a liquidity of 20 Wrapped BNB (almost \$7,180 at the moment of the operation) and 44 trillion of OnlyFans tokens.

After just 6 seconds, on block 8090753, an address swaps 4 million OnlyFans for 0.002 Wrapped BNB (\$0.718). This operation is followed by 11 other swaps---performed by 11 different addresses---for a total buy of $5.1740396*10^{12}$ OnlyFans for 2.67 Wrapped BNB (\$958).
After 2 hours from the creation of the token, at block 8093101  (2021-06-07 03:38:55 PM UTC), the token spammer removes all the liquidity from the liquidity pool, leaving it drained. Since the 12 addresses added Wrapped BNB into the pool by buying OnlyFans, the token spammer collects 22.67 Wrapped BNB and has a profit of 2.67 Wrapped BNB (\$958).

We can formalize these operations in the following way:
\begin{enumerate}
    \item Eve creates a new ERC-20 token $\tau$.
    \item Eve creates a new liquidity pool with pair ($\tau$, $B$), where $B$ is a valuable token, e.g. Wrapped BNB. 
    \item Eve adds liquidity to the liquidity pool. The reserves of the pool are now $(reserve_{\tau}, reserve_B$).
    \item At this point, Eve is the only one that owns token $\tau$. Investors can buy token $\tau$ by swapping their tokens with token $\tau$ in the liquidity pool.
    \item Suppose that Bob buys a few $\tau$ swapping it with $B$. The new reserves of the liquidity pool are $(reserve_{tau} - \delta_{tau}, reserve_B + \delta_B$)
    \item Lastly, Eve removes all the liquidity from the liquidity pool. The net gain of the operations is $\delta_B$ minus the gas fees to execute the transactions.
\end{enumerate}

\mypara{An improved version of the operation.}
The rug pull described above is the simple version of the operation. %, when the attacker never interacts with the liquidity pool until he removes the liquidity. 
However, to attract more investors, the attacker can manipulate some statistics of the liquidity pool. %, such as the number of swaps, the trading volume, or the price.
A well-known market manipulation that the attacker can use is \textit{wash-trading}~\cite{cao2015detecting}.
In this case, the creator of the pool tries to create the impression that the liquidity pool is active, faking the trading volume by repeatedly buying and selling tokens. Similarly, another way that attackers have to drum up the attention of investors is to inflate the price by buying the 1-day token gradually. 
\\
Finally, the attacker can also hedge his gains---eliminating the risk of an unrealized profit while the liquidity pool is still active. The attacker can maintain a reserve of tokens and, when investors start to buy the 1-day token, gradually sell the owned token, starting to take profit from the operation.

Clearly, rug pull operations can harm investors. However, we cannot consider it a "fraud" because the phenomenon is currently not regulated. In Appendix~\ref{sec:are_rug_pulls_frauds} we discuss this subject in depth.
\subsection{Looking for 1-day Rug Pulls}
\label{sec:looking_for_rug_pulls}
We leverage our datasets to identify rug pulls systematically. 
Since we saw a considerable number of 1-day tokens and most of them are created serially, we narrow our investigation to the 332,265 in BSC and 25,180 in Ethereum 1-day tokens with a liquidity pool. Given the duration of these operations, we will refer to them as \textit{1-day rug pulls}.
We analyze all the Events emitted by the liquidity pools, looking for all the pools that emitted only one Mint and one Burn event in which the address that performs the transaction burns at least 99\% of the minted LP-tokens (we don't use 100\% since a small fraction of tokens might be stuck in the wallet due to rounding).
%Similarly, we choose to relax the constraint that the last event emitted by the liquidity pool has to be the Burn. After removing the liquidity, a tiny fraction of the liquidity can remain that can still be used to further swaps.

\subsubsection{Estimating the gains of the operations}
\label{sec:fraud_costs}

The simple operation, where the attacker does not swap in his liquidity pool, can be carried out by performing just four transactions: A transaction that creates the token, one that creates the liquidity pool, one to add the liquidity, and finally, the last transaction to remove the liquidity. These transactions can be performed individually, or they can be aggregated by leveraging a smart contract. Of course, we consider both cases when computing the fees. If the attacker performs swaps on the liquidity pool, we also consider the transaction fees paid for each swap.

To perform our estimation we use the following formula:

\setlength{\abovedisplayskip}{0pt} \setlength{\abovedisplayshortskip}{0pt}

\begin{gather}
\label{for:gain}
base\_gain = \delta_B - \textit{fees} \\
net\_gain = base\_gain - T_{in} + T_{out}  - \textit{fees}_{swap}
\end{gather}
The formula can be split into two components. The first part computes the gain in the case of the simple operation.%, that is the case that we formalized in Sec.~\ref{sec:case_study}.
The second formula takes into account the improved version of the operation, where the creator of the liquidity pool manipulates it by performing swaps operations. In this case, we remove from the gain $T_{in}$, that is the amount of tokens that the manipulator artificially adds to the liquidity. We also add to the gain $T_{out}$, the quantity of tokens that the manipulator removes from the liquidity pool before the final removal of the liquidity ($T_{out}$). Finally, we remove from the gain the fees used to perform the swap operations ($fee_{swap}$).

\subsection{Results}
\label{sec:results}
After processing our data, we discover that an incredibly high number of liquidity pools are actually rug pulls. In BSC, 272,349 out of 332,265 (81.2\%) of the considered liquidity pools have a rug pull pattern, while 21,742 out of 25,180 (86.3\%) in Ethereum. This result shows that attackers use most of the 1-day tokens as disposable to carry out rug pulls.

These operations are arranged by 116,516 different addresses in BSC and 16,539 different addresses in Ethereum. As we can expect from the previous analyses, most of the token spammers that operate in BSC are linked to this kind of activity. Indeed, in BSC, 2,112 out of 4,231 (50\%) token spammers performed at least one rug pull. Instead, in Ethereum, there are only 45 token spammers (0.3\%) that have been involved in this activity.
We find 115 addresses that perform more than 100 rug pulls in BSC, accounting for 19.1\% of the operations, with the most active performing 16,102 operations. 
Instead, in Ethereum, we find only one address performing more than 100.
Interestingly, combining the information in the BSC and Ethereum dataset, we find a token spammer that operated on both blockchains with the same address~\footnote{0x87605612492c74bA0037fFaef676c0f3f6958918}. He performs five rug pulls on Ethereum and three on BSC.

%\mypara{Exit scams involve hundred thousand of swaps.}
Looking at the liquidity pools, we find that BNB (97.8\% of the cases) is the token paired the most with the 1-day token. It is followed by USDT (0.67\%) and BUSD (0.15\%), two stablecoins pegged to the USD. Instead, Wrapped Ether is paired with all the 1-day tokens in almost all the liquidity pools with a rug pull in Ethereum. As the next step, we want to estimate the number of users that fall prey to such activities.
To do so, we exclude the addresses that swap into liquidity pools they have created themselves from this analysis. 
We collect 251,250 different addresses in BSC and 57,552 in Ethereum that interact with at least one liquidity pool with a rug pull pattern. These addresses performed 2,903,022 swaps on the considered liquidity pools in BSC and 317,257 in Ethereum.

We divide the swaps into buy (1-day token) and sell operations.
As we can expect, given the anatomy of the 1-day rug pull, we find that most of the operations are buy operations. More in detail, in BSC 2,286,056 (78.7\%) are buy operations and 616,966 (21.3\%) sell operations. In Ethereum, we find a very similar pattern, with 254,061 (80.1\%) buy operations and 63,196 (19.9\%) sell operations.

As final metric, we compute the average value of the swaps performed by the users. The average amount of swaps is almost identical for buy and sell operations on both the blockchains, with 0.01 BNB for BSC and 0.19 ETH for Ethereum's liquidity pools. Interestingly, we notice a considerable difference in the average swap amount between the two blockchains. Indeed, the average swap is approximately \$3 on BSC and \$360 on Ethereum.

% \begin{table}
%    \centering
%    \caption{%
%       Market manipulations in liquidity pools where an exit scam occurs, %divided by type of operation.
%    }\label{tab:manipulations}
%    \begin{tabular}{@{\extracolsep{2pt}}l r r}
%        \toprule
%        Operation & BSC  & ETH \\
%        \midrule
%        Wash-trading & 19,833  (42.3\%) &  1,031 (28.7\%) \\ 
%        Pump  & 26,463 (57.7\%) & 2,689 (71.3\%) \\ 
        %Wash-trading & 19,833  (33.3\%) &  1,081 (21.7\%) \\ 
        %Pump  & 26,463 (44.3\%) & 2,689 (53.4\%) \\ 
        %Hedge & 13,324 (22.4\%)   & 1,210 (24.2\%)  \\ 
%        \midrule
        %\# of liquidity pools & 59,670 & 4,980 \\ 
%        \# of liquidity pools & 46,296 & 3,770 \\ 
 
%    \end{tabular}
%\end{table}

\subsubsection{The gains}
\label{sec:gains}
Before computing the gains of the attackers, we calculate the average price an attacker has to invest to arrange the operation.
If the attacker does not perform any swap into the liquidity pool, the cost of the operation is on average 0.03 BNB in the case of BSC and 0.2 ETH for the Ethereum blockchain. 
Thus, the investment needed to perform such operations is low, even if it could vary substantially when the blockchains are overloaded. For instance, we found some rug pulls that reached the cost of 1.1 BNB or even 3.3 ETH. 
The base cost to arrange the operation is interesting because it represents a bound to the loss the attackers have to afford for each operation.

We leverage our datasets to compute the gain of the operation using the formula~\ref{for:gain} described in Sec.~\ref{sec:fraud_costs}. We describe the 266,340 operations on BSC and the 21,594 on Ethereum in terms of successful and unsuccessful operations based on the operation's net gain. In particular, we consider an operation successful if the net gain is strictly positive.

\textbf{Successful operations}.
Among the liquidity pools with a rug pull pattern, there are 104,404 (39.1\%) operations in BSC and 13,368 (61.9\%) in Ethereum closed with a profit for the attacker.
A possible reason for the higher success rate of the rug pull on Ethereum could be that, as we saw, on average, users tend to invest more money. Indeed, on average, attracting only one investor is enough to cover the operation's cost.
To investigate what can affect the gains, we combine information on gains with those of the manipulations.
%Tab~\ref{tab:manipulations} shows the results.
When the creator of the liquidity pool does not perform any kind of manipulation, the net gain is, on average 0.11 BNB in BSC  and 1.34 ETH in Ethereum.
Operations carried out on liquidity pools that suffer wash-trading activity have an average gain of 0.25 BNB in BSC and 12 ETH in Ethereum, which is considerably higher than the previous case.
Instead, we notice a negligible increase in gains in the case of pump operations with respect to the gains obtained by the liquidity pools without manipulation.
Moreover, we find that both kinds of manipulation have no impact on the success rate.
This show that operations that have wash trading are generally more profitable. However, the attacker has to perform several swaps, increasing its cost and loss in case of an unsuccessful operation.

\textbf{Unsuccessful operations.}
There are 161,936 (60.9\%) liquidity pools in BSC and 8,226 (38.1\%) in Ethereum, for which the attacker does not cover the transaction fees with the operations.
For the 14\% (21,122) of these liquidity pools of BSC and the 20\% (1,506) of Ethereum, we notice that the operations were unsuccessful because nobody swapped into the liquidity pools. Considering the results we obtained, we can conjecture that the aim of the attackers is not to be successful every time but to arrange rug pulls serially and take profit in the long run. Indeed,  the loss of an unsuccessful operation is minimal, and a streak of operations closed in loss can be covered with a single profitable operation.

\textbf{Financial cost of 1-day rug pulls and comparison with other blockchain phenomena.}
In our study, we find that the number of 1-day rug pulls (21,594) and attackers (16,439) in Ethereum is significantly lower than in BSC (266,340 operations carried out by 117,110 rug pullers).
Nonetheless, the total gain of Ethereum operations, around \$150 million, is remarkably higher than the gains of BSC operations, that amount to \$91 million. 
Moreover, the same trend holds when considering the volume of rug pull operations, which we define as the total value of BNB and ETH swapped. Here we find that Ethereum has a volume of  \$772.5 million against the \$243.5 million of BSC.
To gain insight into the magnitude of 1-day rug pull operations, we compare our metrics with popular blockchain shenanigans, like MEV and front-running. Tab.~\ref{tab:comparison_frauds} in Appendix reports more relevant metrics collected from related works about operations carried out in Ethereum.
As we can see, 1-day rug pull is the second type of operation by profit, generating slightly lower gains than Sandwich Attacks (\$174.34 million in accordance with Qin et al.~\cite{qin2022quantifying}).
Particularly interesting is the number of addresses that perform the operations. Indeed, in Ethereum, the number of attackers that performed  1-day rug pulls is almost five times the number of the Sandwich Attackers (the fraud with the higher number of attackers in our comparison). We believe the operations are performed by a large number of addresses due to their ease of execution. The reported numbers highlight that 1-day rug pulls is a significant phenomenon in the DeFi ecosystems that involve hundreds of thousands of malicious actors and move more than 1 billion USD.

\subsubsection{A longitudinal view}
\label{sec:longitudinal}
\begin{figure*}
    \centerline{\includegraphics[width=1.\textwidth]{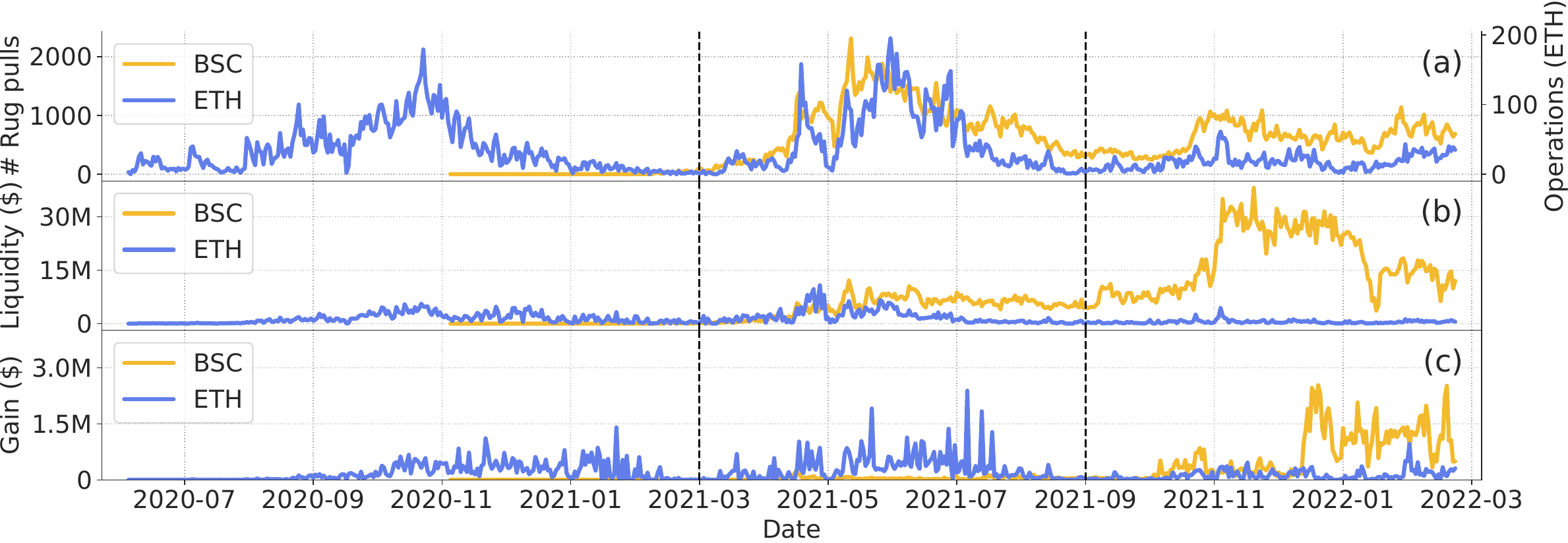}}
\caption{The figure shows the number of rug pull operations (a), the initial liquidity added to each pool (b), and the gain for each operation over time. All the metrics are aggregated daily. The dashed vertical lines divide the three phases we identify.}
  \label{fig:longitudinal}
\end{figure*}

Fig.~\ref{fig:longitudinal} provides a longitudinal view of the daily number of rug pull operations (Fig.~\ref{fig:longitudinal} a), the liquidity added (Fig.~\ref{fig:longitudinal} b) and the gains (Fig.~\ref{fig:longitudinal} c). 
Analyzing the trends of the chart, we identify three different phases, divided by the black dashed lines in the figure.
%The chart shows three interesting phases, divided in the figure by the black dashed lines:
%Looking overall at the three phases we can see an interesting trend. 
In the first phase, we find the first spike of 1-day rug pulls in Ethereum. In the second phase, rug pulls start to increase in the BSC. However, the Ethereum gains are generally higher for the same invested liquidity. Finally, in the third phase, we see that BSC surpasses Ethereum in terms of liquidity added, number of operations, and gains.
In the following, we describe in detail the three phases:\\
\textbf{Phase 1: DeFi Summer.}
The first phase took place approximately from June 2020 to March 2021.
At the beginning of this phase, we see an increase in the daily number of rug pull operations in Ethereum, with a peak of 179 daily operations in October 2020. Then, the number of operations steadily decrease until March 2021. 
We believe that the increase in the number of operations was bootstrapped by a phenomenon known in the crypto-community as DeFi Summer 2020~\cite{maouchi2022understanding}.
During this period, DeFi became extremely popular, and, as a result, the market capitalization and prices of several tokens soared~\cite{defisummer}. 
This interest in DeFi attracted new users looking for investment opportunities, which may have triggered the increase in rug pull operations.
Fig.~\ref{fig:longitudinal} (b) shows that there is a significant amount of liquidity invested in these operations, on average \$37,941 (44 ETH), with an average gain of \$5,969 (5.65 ETH) (Fig.~\ref{fig:longitudinal}). 
Note that this phase involves only the Ethereum blockchain because the BSC was released in September 2020 and was not very popular yet. \\
\textbf{Phase 2: Altcoin season.}
Fig.~\ref{fig:longitudinal} (a) shows a second spike in the number of rug pulls from March 2021 to September 2021. 
In this case, the spike involves both Ethereum and the BSC, which reach a maximum peak of 195 and 2,309 daily operations. It is interesting to notice that the number of operations over time follows the same trend for Ethereum and BSC. For this reason, we believe an exogenous event caused this spike. Analyzing the events of that period, we believe this rise in the number of operations may be a so-called \textit{Altcoin Season}.
An Altcoin Season is a period in which Altcoins\footnote{Altcoins~\cite{haferkorn2014seasonality} is a combination of the two words "alternative" and "coin". The term is used to indicate all cryptocurrencies except Bitcoin.} perform better than Bitcoin, significantly increasing their value.
Previous study~\cite{kulal2021followness} shows that an Alt Season is marked by a drop of an indicator called \textit{Bitcoin dominance}.
%it is possible to detect an Alt Season by monitoring the decrease of an indicator called \textit{Bitcoin dominance}. 
This indicator measures the ratio between the market capitalization of Bitcoin to the total market capitalization of the entire cryptocurrency market.
According to Coinmarketcap~\cite{coinmarketcap}, in this period, the  Bitcoin dominance decreased from 69\% of January 2021 to 39\% in May 2021.
This market phase is frequently characterized by "Fear of missing out" (FOMO)~\cite{baur2018asymmetric}, which makes investors more inclined to buy riskier tokens.
For this reason, we believe investors have flocked to AMM markets to buy tokens, and rug pull operations skyrocketed.
Fig.~\ref{fig:longitudinal} (b) shows that the liquidity invested in these operations is higher for Ethereum, with an average of \$39,625 (50 ETH) against the \$5,624 (15 BNB) of BSC.
Operations in Ethereum are also way more profitable, with an average gain of \$5,836 (6.3 ETH) against the \$48.4 (0.12 BNB) of BSC operations. \\
\textbf{Phase 3: The overtaking of the BSC.}
The last phase goes from  October 2021 to March 2022. 
In this phase, we find an interesting twist, as BSC surpasses Ethereum in terms of liquidity added and gains of rug pull operations.
Indeed, in this phase, rug pulls in BSC have significantly more liquidity invested than in the past (56.9 BNB on average vs. 15.3 BNB of the previous phase) and higher gains (2.26 BNB on average vs. 0.12 BNB of the previous phase).
For this reason, we can see in Fig.~\ref{fig:longitudinal} that the total daily invested liquidity and gains in BSC are significantly higher than Ethereum and reached more than one million USD.
%Since we did not find any external cause for this increase, it is possible that BSC replaced Ethereum as a platform of choice for rug pulls. 
In Sec.~\ref{sec:discussion}, we explore some possible reasons for this increase. 

%The number of daily rug pulls in Ethereum and BSC follow the same trend, but in terms of raw number, the BSC shows ten times more operations than Ethereum. What is interesting in this phase is the significant increase in invested liquidity and gains of exit scams in BSC.

\subsubsection{Tokens' names}
\label{sec:tokens_names}
                  
To further deepen our analysis of rug pulls, we focus on the names used in the operations.
Analyzing the rug pulls, we notice several tokens with the same name in BSC and Ethereum. We find that of the 272,349 tokens involved in the operations in BSC and 21,742 in Ethereum there are only 157,864 (57.9\%) and 18,801 (86.4\%) unique names. 
Thus, we attempt to cluster the 1-day tokens into categories and enumerate them.
Table~\ref{tab:fakes} in the Appendix shows the most used names and the number of occurrences for each of them. 

As a first category, we explore clones---tokens with the same name as an existing (and more popular) cryptocurrency. 
To systematically search for these cases, we use as an authoritative source the CoinGecko APIs~\cite{coingeckoapi}. Leveraging them, we retrieve the names and the addresses of all tokens created and verified with the indexer service on the BSC and Ethereum.
At the end of the process, we build a list of 5,325 tokens for BSC, and 5,172 tokens for Ethereum. We complement this list by adding popular variations for some tokens' names (\eg we also considered ADA as a possible name for the Cardano token).
Using our list, we discover 22,002 cloned tokens in BSC and 1,781 in Ethereum. The most cloned tokens in BSC are Berryswap (370), Shiba Inu (191), and SafeMoon (158).

The second category we explore is the one of tokens that attempt to impersonate companies or websites. In this case, to obtain a list of possible target companies, we retrieve the name of the companies of the Standard and Poor's 500 (S\&P 500) stock market index.
Instead, for the websites, we extract from the Alexa ranking~\footnote{Data retrieved 2022-04-26} the name of the top-ranked 200 websites.
Using in conjunction these two lists, we find 4,638 tokens of this category in BSC and only 95 in Ethereum. The companies and websites that are present the most are Pornhub (1,023), Spacex(419), Onlyfans (398), Oracle (319), and Amazon (270). 

We find several names that contain popular meme-related words like "Doge", "Inu" or "Shiba". 
This is not surprising, since meme tokens are very popular after the events that involved the "meme stocks" of GameStop (GME) and AMC Entertainment (AMC) in late 2020~\cite{morgia2021doge}. 
Luckily, CoinMarketCap and CoinGecko offer a categorization of the tokens that also contain the "meme" category. We leverage these lists to extract the most frequent words and search for them into the tokens involved in rug pulls. We find a huge amount of tokens of this category: 54,229 in BSC and 4,835 in Ethereum.

As the last category of our investigation, we look for DeFi services (\eg Deriswap, Shibaswap, and Eco Finance).
In this case, we simply search for tokens containing the "swap", "defi" and "finance" keywords. With this approach, we find for this category 25,524 tokens in BSC and 3,751 in Ethereum. 

With our simple categorization, we covered the names of 39\% of the 1-day tokens on the BSC and 48\% on Ethereum. Even if we were not able to categorize all the tokens, we get some insights on how attackers pick the name to arrange their operations. In particular, we note a strong trend in choosing tokens' names related to the meme category and leveraging the name of popular cryptocurrencies, services, and companies. 

\section{Sniper Bots 2.0}
\label{sec:sniper_bots}
\begin{figure}
    \centerline{\includegraphics[width=0.50\textwidth]{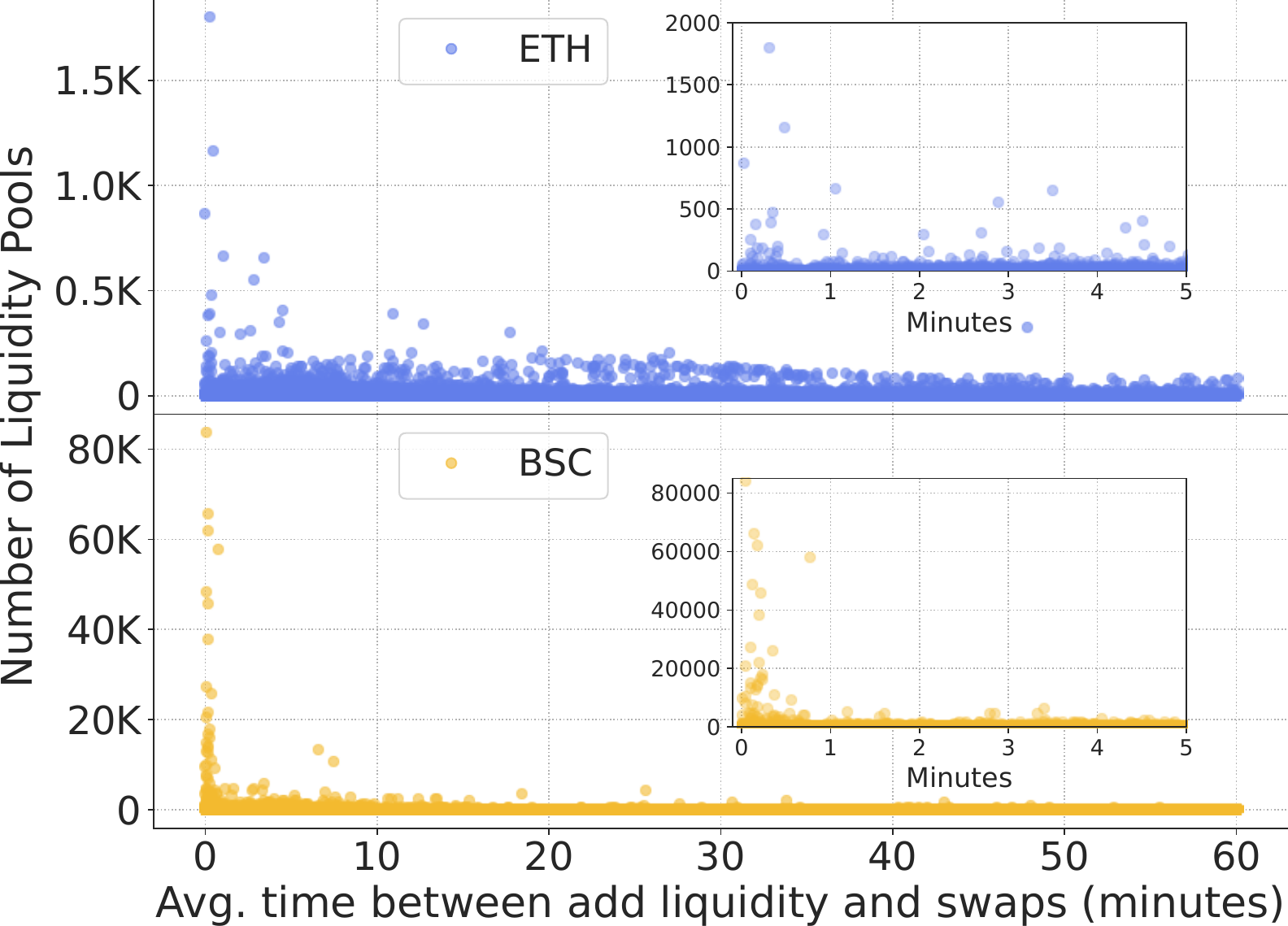}}
\caption{
Scatter plot of the number of liquidity pools with a 1-day rug pull pattern where the address swapped and the average delay from
the pool creation.}
%\caption{Each data point represents an address that swaps inside liquidity pools with a rug pull pattern. On the \textit{y axis}, we represent the number of different liquidity pools where the address swaps. On the \textit{x axis} we show the average time interval between the first time the liquidity is added to the liquidity pool and the swaps operations of the address.}
  \label{fig:snipers}
\end{figure}
We find that a large fraction of rug pulls are successful, even if they are zero-effort operations, without fake tokens or wash trading. Since these kinds of operations are very quick and simple, it is still unclear how they can be profitable. We analyze the operations carried out inside rug pulls more in-depth and discover that their success may be due to the activity of a particular class of trading bots, called Sniper Bots.

Sniper bots are automated bots that monitor time-bound activities and perform an action before or after anyone else.
An example of sniper bot are ``Scalping Bots'', bots that monitor the availability of target products from a website and buy them as soon as they are available (\eg Nvidia GPUs)~\cite{brock2021scalping}.

With the birth of and the widespread adoption of AMMs, a new kind of sniper bot has been developed, which we define \textit{Sniper Bots 2.0}.
These kinds of sniper bots are programs that buy tokens on liquidity pools as soon as they are listed. %\mlnote{Thus, the basic idea is to ``snipe'' a new token by buying it before anyone else at its initial price.}
To do so in the fastest way, sniper bots can leverage the mempool--- the list of transactions not yet inserted in blockchain blocks. %\mlnote{In this aspect, sniper bots are similar to front-running bots ~\cite{daian2019flash}, that instead leverage the mempool to profit on swap operations by anticipating them.}
We find examples of these bots distributed for free on Github~\cite{sniper1,sniper2,sniper3} and for a price at several other websites~\cite{sniper_website_1,sniper_website_2}.
Analyzing the code, we can infer how they work.
As a first step, the sniper bot must search for newly listed tokens. The fastest implementation scans the mempool looking for transactions whose byte-code indicates that they are adding liquidity to a brand new liquidity pool.
Another possibility is that the sniper bot waits for the token to be listed on services like BscScan or Etherscan.
Then, the bot sends a swap transaction to buy the token, and if the gas price is properly adjusted, it is executed in the same block (but immediately after) of the transaction that adds the liquidity. Sniper bots typically execute only the buy operation. The user then can freely decide when to sell the token and make a profit. However, we also found some variants that automatically sell the token when the price reaches a pre-defined goal.
%Note that this activity is not illegal.

\subsection{Identifying Sniper Bots} 
We conjecture that one of the reasons for the profitability of rug pulls operations are sniper bots that buy tokens from every liquidity pool indiscriminately. Thus, we can consider the liquidity pools involved in rug pulls as ``honey pots'' to detect sniper bots. To verify our intuition, we focus on addresses that swapped inside liquidity pools with a rug pull
Fig.~\ref{fig:snipers} shows the phenomenon: Every dot is an address, and its position indicates the number of different liquidity pools where the address swapped and the average delay from the pool creation.
The figure shows a few addresses that swap in thousand of liquidity pools almost immediately after their creation.
Since these addresses perform these operations serially and incredibly fast, we believe they must be sniper bots. 
We set up two conservative thresholds to identify evidence of addresses used by sniper bots.

For BSC, we consider all the addresses that swap on average with a delay smaller than five blocks (15 seconds) and that swap in at least 100 different liquidity pools.
We flag 130 addresses as possible sniper bots. These addresses represent only 0.03\% of all the addresses that swap inside liquidity pools involved in rug pulls. What is impressive is that they swap in 235,777 liquidity pools, representing 68.7\% of all the liquidity pools with a rug pull. 
Moreover, these addresses also  perform an impressive number of swaps: 2,691,173, that account for 24\% of all the swaps performed in liquidity pools with a rug pull.
We find that 31\% of these swaps are performed in the same block where the liquidity is added for the first time in the liquidity pool. In these cases, we can confirm that the sniper bots scanned the mempool to swap in the same block where the liquidity is added. However, we also find sniper bots that perform the swap operations a few blocks after the liquidity is created. 

We find sniper bots to be less present in Ethereum. Also, in this case, we pick two thresholds and consider all the addresses that swap on average with a distance lower than three blocks (45 seconds) and that swap in at least 10 liquidity pools.
% that are only the 0.1\% of all Ethereum addresses that swap in liquidity pools with a rug pull pattern.
We find 64 possible sniper bots that swap in 30\% of all the liquidity pools and perform a much smaller fraction of swaps with respect to BSC sniper bots (3.5\% of the total). However, interestingly, a higher percentage of swaps are performed in the same block where the liquidity is added in the liquidity pools (60\%).

\section{1-day Rug Pull Mitigation}

Our study highlights that the 1-day rug pulls have some distinctive features.
In the following, we propose some metrics that stem from the lessons learned from our analysis that may be useful to build a detection system.
\begin{itemize}
     \item \textbf{Token lifetime:}
     This metric measures the time that elapses since the creation of the token. Indeed, we find that 1-day rug pull operations are performed in a very short timeframe (\S \ref{sec:looking_for_rug_pulls}). 
    \item \textbf{Distribution of the liquidity:} This metric tracks the distribution of the LP-tokens.
    In 1-day rug pulls, the liquidity pool creator owns all the liquidity (\S \ref{sec:case_study}). Thus, it should be considered extremely risky when a single address owns most of the liquidity.
    \item \textbf{Address rug pull records:} This metric tracks addresses that performed a rug pull operation to add them to a list of potential malicious addresses.
    Indeed, we find that some addresses perform rug pulls multiple times. (\S \ref{sec:serial}).
    \item \textbf{Deceptive token name:} This metric measures the similarity between the name of tokens contained in the liquidity pools and popular existing tokens or companies. We find that attackers often deceive investors by exploiting the name of the token. (\S \ref{sec:tokens_names}).
\end{itemize}
%The system evaluates the proposed metrics on each liquidity pool. Then, it combines the metrics using a weighting schema to define its the risk level. The main risk in introducing the proposed system is to flag recently created benign tokens.
%However, this can be mitigated by properly tuning the weight associated with each metric. For instance, giving some features (\eg "Distribution of the liquidity") more weight than the "Token lifetime".
%Of course, an attacker aware of this system can try to deceive the metrics by putting more effort into carrying out the operations (\eg using different addresses or creating the token in advance). 
An attacker aware of these metrics can try to evade the detection by putting more effort into carrying out the operations (\eg using different addresses or creating the token in advance).
Nonetheless, a distinctive characteristic of 1-day rug pulls is that they are easy to execute and require low effort by the attacker. Thus, we believe the proposed metrics could be sufficient to discourage this operation.
Moreover, new metrics and more sophisticated techniques can be developed to identify attackers trying to circumvent the detection. For example, it is possible to follow the money flow between addresses associating different addresses to the same attacker.

We believe that AMMs are interested in leveraging the proposed metrics to build a detection system. Indeed, some have already put effort into this direction. For instance, PancakeSwap recently included in its interface a service called HashDit~\cite{hashdit}, which provides a risk level in investing in a liquidity pool.
HashDit is a Token Contract Scanning service, that estimates the risk of a token by analyzing the code of its smart contract~\cite{hashditbinance}. We believe the proposed metrics can enhance this and other existing services by adding insightful information.

\section{Related Work}
\label{sec:sota}

\mypara{Tokens identification.} 
In previous work, there are mainly two token identification approaches: behavior-based and interface-based. The behavior-based method assumes that a token contract maps addresses to the number of tokens owned and contains a function to transfer tokens. Chen et al.~\cite{chen2019tokenscope} follow this approach, analyzing the EVM execution path to find smart contracts data structures that indicate the bookkeeping of a token. The interface-based approach, the technique we take in this work, aims to find tokens that conform to specific interfaces (e.g., the ERC20 interface). This method involves discovering the implemented functions within the smart contract bytecode. Several works use this approach~\cite{di2021identification, victor2019measuring, chen2020traveling}. Frowis et al.~\cite{frowis2019detecting} proved that the interface-based technique could detect 99\% of the tokens in their ground truth dataset.

\mypara{Liquidity pool scams.} Xia et al.~\cite{xia2021trade} characterize scam tokens on Ethereum. First, they leverage CoinMarketCap~\cite{coinmarketcap} to obtain a ground truth of official and scam tokens. They used The Graph~\cite{thegraph} to obtain 21,778 tokens and 25,131 liquidity pools from May 2020 to December 2020. A guilt-by-association heuristic is adopted to enlarge the dataset, subsequently used to train a machine learning model. More than 11,182 fraudulent tokens were discovered after they ran their classifier on the expanded dataset. Mazorra et al.~\cite{mazorra2022not} extended Xia et al. dataset by including Uniswap data until 3 September 2021, discovering an additional 18 thousand scam tokens. They provide three categories for rug pulls: simple, sale, and trap-door. Then, they found that more than 97.7\% of the tokens labeled as scams are involved in rug pulls.

\mypara{Rug pull mitigation.} Rug pulls are a very recent issue, and to the best of our knowledge there is no actual solution to prevent them.
However, there is a new proposed standard and some protocols that can help to mitigate the problem.
To counter the theft of tokens, Wang et al.~\cite{wang2022erc} proposed a new token standard called ERC-20R. With this standard, a transaction is reversible for a short time (dispute period) after it has been performed. During this period, the sender can request to freeze the disputed asset to a set of decentralized judges. If judges agree to lock the disputed asset, it starts another period of time in which the sender can convince judges to revert the transaction.
Instead, liquidity locker protocol (\eg Unicrypt~\cite{unicrypt}) allows locking LP-tokens inside smart contracts for a given amount of time. This solution assures that the liquidity cannot be removed from the pool until the timer expires, making rug pull impossible. Of course, this solution does not prevent rug pulls after the time expires or dumping one of the tokens in the liquidity pool.

\section{Discussion}
\label{sec:discussion}

\mypara{What is the impact of not collecting all the internal transactions?}
Unlike other works~\cite{victor2019measuring, chen2020traveling}, we do not collect all smart contracts generated by internal transactions.
We collect smart contracts created directly by EOAs, and expand our dataset by adding contracts that emitted at least one Transfer Event.
This approach could lead to the loss of a small percentage of tokens. We can perform a rough estimation of the ERC-20 token we miss by comparing the number of tokens we retrieved with the number of tokens retrieved by Chen et al.~\cite{chen2020traveling} at the same block height.
%(see Tab.~\ref{} in Appendix).
Our approach retrieves 146,928 tokens instead of 165,955, approximately 12\% less. However, it is important to note that, by design, our approach misses only tokens that are never used, traded, or transferred. So, the missing tokens do not represent interesting cases for our study.

\mypara{Why does it appear that rug pulls and token spammers are more frequent in BSC than in Ethereum?}
From a technical point of view, rug pulls work the same way in the two blockchains. Indeed, since BSC is EVM compliant and PancakeSwap is a fork of Uniswap, the same smart contract can be used on both blockchains. However, the cost of the operation is significantly different.
As we saw in Sec.~\ref{sec:fraud_costs}, performing a rug pull in BSC is cheaper (on average \$10.5 with peaks of \$600) than in Ethereum (on average \$400 with peaks of over \$2,000). These costs represent a fixed cost for the attacker, and going even or gaining money may be more difficult in Ethereum versus BSC.

%\mypara{Is cost the only factor for the high number of rug pulls in the BSC?}
\mypara{Are cost-efficient blockchains vulnerable to 1-day rug pulls?}
As discussed, one of the possible reasons for the prevalence of rug pulls on BSC is the low transaction cost. This could suggest that cost-efficient blockchains are more vulnerable to 1-day rug pulls. However, to confirm this hypothesis, it is necessary to examine whether the phenomenon is common in blockchains with costs similar to the BSC. 

Considering our case study of BSC, we believe the low cost of transactions is not the only reason for the high number of rug pulls.
In particular, BSC provides one of the first DeFi ecosystems that is cheaper and faster than Ethereum. It quickly became very popular.
Moreover, thanks to EVM compatibility, many no-code tools, libraries, and smart contracts already developed for Ethereum can also be used on BSC. This allows the deployment of smart contracts and the creation of new tokens with limited technical capabilities. 
Thus, the high number of potential victims, the little technical challenge, and the cost-efficiency made the BSC fertile ground for malicious actors to carry out 1-day rug pulls.
Even though the low cost can facilitate rug pulls, increasing the costs of blockchains is not a real solution. 
Instead, a possibility is to shift the focus to DEXes's protocol and smart contracts for token creation.
In particular, it could be possible to design more secure smart contracts to handle tokens (\eg ERC-20R) or AMM protocols with policies that disincentive rug pull operations.

\mypara{Can different users coordinate to carry out the same operation, or can a user use multiple addresses?}
In this work, we considered each address belonging to a single different user, and we assumed there is no coordination among addresses. Nonetheless, a user may change the address he uses to perform each rug pull. It is also possible that a group of users coordinate to carry out the operation. For example, a user can create a liquidity pool while others perform wash trading.
A possible approach to detect this malicious behavior is to gather all the transactions among the allegedly involved addresses and look for malicious patterns or communities (\eg using graph analysis). In this work, we do not perform this analysis, but we plan to explore more sophisticated rug pulls as an extension of this work.

\section{Ethical considerations}
\label{sec:ethical_considerations}

In this paper, we examined 3 billion transactions from Ethereum and the BSC. We focused our research on the addresses that create tokens and how they use them. All data we retrieved is publicly available, and EOA addresses are pseudo-anonymous.  We never attempted to deanonymize the addresses or violate their privacy during this work. Consequently, and in accordance with our IRB's policies, we did not require express approval to conduct our analysis.

\section{Conclusion and Future Work}
\label{sec:conclusion}

In this work, we conduct a thorough investigation of the tokens and the liquidity pools of the BNB Smart Chain and Ethereum.
We studied the lifetime of the tokens and their creators. We
discovered two very interesting metrics: 60\% of the total tokens of both blockchains do not survive their first day (1-day token), and a tiny fraction of addresses (1\% of addresses), which we called token spammers, created more than 20\% of the tokens. 
We explore the correlation between token spammers and 1-day tokens, and we found that token spammers strongly impact the existence of 1-day tokens. 

More interestingly, we find that token spammers use 1-day tokens as disposable tokens to arrange rug pulls, exploiting the mechanism of liquidity pools. 
We selected from our dataset all the liquidity pools that show evidence of a rug pull and dissect the operations, analyzing them from several perspectives. 
Finally, we introduce the sniper bot, trading bot that aims to buy tokens at their listing price. However, they unwillingly became victims of the rug pulls because of their mechanism.

As future work, we believe it is interesting to further refine our results by including addresses that cooperate to perpetrate rug pulls in the analysis. It could be possible to uncover other malicious and more sophisticated patterns.  
As discussed in Sec.~\ref{sec:discussion}, cost-efficient blockchains could be more exposed to the 1-day rug pulls. Thus, it is interesting to extend our analysis to blockchains with transaction costs comparable to BSC (\eg Algorand~\cite{gilad2017algorand}).
Finally, another promising direction is further exploring sniper bots to provide a more detailed analysis of their typologies and operations. 

\section*{Acknowledgments}
This work was partially supported by project SERICS (PE00000014) under the MUR National Recovery and Resilience Plan funded by the European Union - NextGenerationEU.
This work was partially supported  by the MIUR under grant ``Dipartimenti di eccellenza 2018-2022" of the Department of Computer Science of Sapienza University.

\bibliographystyle{plain}
\bibliography{biblio}

\clearpage
\appendix

\begin{table*}[t]
   \caption{%
       Comparison with other blockchain operations. We report only on frauds performed on Ethereum since our work is the first to analyze the Binance Smart Chain.}
  \label{tab:comparison_frauds}
  \begin{tabular}{l r r r r l l}
      \toprule
      Operation & Tot Gain (\$) & \# Addresses & \# Operations & Blockchain & From & To\\
      \midrule
      %BEV Arbitrage & \cite{qin2022quantifying} & 277.02M & 8,769 & 1,151,448 & Ethereum & 2018-12-1 & 2021-8-5 \\
      \cite{gao2020tracking}~Counterfeit Tokens   & 17.35M & 364 & 573 & Ethereum & 2015-07-30 & 2020-03-18 \\
      \cite{torres2021frontrunner}~Displacement   & 4.1M & 74 & 2,983 & Ethereum & 2015-07-30 & 2020-11-21 \\
      \cite{qin2022quantifying}~Fixed Spread Liquidations   & 89.18M & 2,724 & 31,057 & Ethereum & 2018-12-01 & 2021-08-05 \\
      \cite{torres2019art}~Honeypots   &  90K & 53 & 690 & Ethereum & 2015-08-07 & 2018-10-12 \\
      \cite{torres2021frontrunner}~Insertion   & 13.9M & 1,975 & 196,691 & Ethereum & 2015-07-30 & 2020-11-21 \\
      \cite{qin2022quantifying}~Sandwich Attacks   & 174.34M & 3,488 & 750,529 & Ethereum & 2018-12-01 & 2021-08-05 \\
      \cite{chen2021sadponzi}~Smart Ponzi   & 17.70M & 444 & 835 & Ethereum & 2015-08-01 & 2020-05-20 \\
      \cite{torres2021frontrunner}~Suppression   & 1.03M & 128 & 50 & Ethereum & 2015-07-30 & 2020-11-21 \\
      %One-hour & 452,708 (42.0\%) & 115,616 (47\%)\\
      \midrule
      1-day rug pulls & 148.93M & 16,439 & 21,594 & Ethereum & 2015-07-30 & 2022-03-07\\
      1-day rug pulls & 90.78M & 117,110 & 266,340 & BSC & 2020-04-20 & 2022-03-07\\
       \end{tabular}
\end{table*}

\section{Functions and events of the ERC-20 and BEP-20 standards}

\begin{table}[h]
\small
    \centering
    \caption{%
       Functions and events of the ERC-20 (Ethereum) and BEP-20 (Binance Smart Chain) standard interface. We report in yellow the methods that are optional in the ERC-20 interface and in red the only method that is optional in both interfaces.
    }\label{tab:erc20signature}
    \begin{tabular}{l r r}
        \toprule
        Function & Signature\\
        \midrule
        \cellcolor{yellow!50}name() &  06fdde03 \\
        \cellcolor{red!50}symbol() &  95d89b41 \\
        \cellcolor{yellow!50}decimals() &  313ce567 \\
        totalSupply() & 18160ddd \\
        balanceOf(address) & 70a08231 \\
        transfer(address,uint256) & a9059cbb \\     
        transferFrom(address,address,uint256) & 23b872dd \\     
        approve(address,uint256) & 095ea7b3 \\ 
        allowance(address,address) & dd62ed3e \\ 
        \midrule
        Event & Signature \\
        \midrule
        Transfer(address,address,uint256) & ddf252ad \\ 
        Approval(address,address,uint256) & 095ea7b3 \\ 
    \end{tabular}
\end{table}

%\section{An example of a 1-day rug pull operation}
%\label{sec:anatomy_rug_pull}
%We focus on \textit{OnlyFans}\footnote{0xe8b6f08841d668605343A63144D76ff2dE9A1199}, a token created by the top token spammer on block 8090747 
%(2021-06-07 01:40:34 PM UTC) by issuing a contract creation transaction. This token has a supply of $10^{13}$ and its symbol is the unicode \textit{U+128139}, the emoji of a kiss mark, followed by the string "OnlyFans". On block 8090751 (2021-06-07 01:40:46 PM UTC), after 4 blocks from its creation, the token spammer creates a liquidity pool that contains the pair (OnlyFans, WrappedBNB) and adds a liquidity of 20 Wrapped BNB (almost \$7,180 at the moment of the operation) and 44 trillion of OnlyFans tokens.

%After just 6 seconds, on block 8090753 (2021-06-07 01:40:52 PM UTC), an address swaps 4 million OnlyFans for 0.002 Wrapped BNB (\$0.718). This operation is followed by other 11 swaps---performed by 11 different addresses---for a total buy of $5.1740396*10^{12}$ OnlyFans for 2.67 Wrapped BNB (\$958).
%After 2 hours from the creation of the token, at block 8093101  (2021-06-07 03:38:55 PM UTC), the token spammer removes all the liquidity from the liquidity pool leaving it drained. Since the 12 addresses added Wrapped BNB into the pool by buying OnlyFans, the token spammer collects 22.67 Wrapped BNB and has a profit of 2.67 Wrapped BNB (\$958).

\section{Are 1-day rug pulls frauds? }
\label{sec:are_rug_pulls_frauds}

%The kind of operations we analyze in our work are very different from well-known rug pulls like Squid Game~\cite{squidgame} or Luna Yield~\cite{lunayield}. 
1-day rug pulls are very different from more notorious rug pulls like Squid Game~\cite{squidgame} or Luna Yield~\cite{lunayield}. 
Indeed, these operations lasted weeks or months, and their perpetrator exploited extensive marketing campaigns and misleading advertising to deceive users into investing in their tokens.
In the case of Squid Game, the scammer created a token in the BSC following the success of the homonym Netflix television series~\cite{squidsuccess}.
%The token was listed on PancakeSwap in October 21, 2021, and was worth a few cents at first. 
%The price of  from its creation in October 2021. 
Due to the extensive marketing campaign promoting the token as official on social media platforms such as Twitter and Telegram, its value skyrocketed from a few cents to over \$2,856 in less than a week~\cite{squidgamecointelegraph}.
Then, the scammer removed nearly all of the liquidity from the pool (\$3.3 million), causing the token's value to plummet to near zero~\cite{squidgame}. 
In our paper, we study 1-day operations that aim to make a profit with the least possible effort in a short time frame.
For this reason, it is unlikely that they leverage sophisticated marketing campaigns to lure investors, like in the case of Squid Game. 
However, some 1-day rug pull operations use other kinds of deceptive tactics.
The first uses token names identical or slightly different from well-known companies or popular tokens. As we saw in Sec.~\ref{sec:tokens_names}, this case involves 8.7\% of Ethereum rug pulls, and 10\% of BSC rug pulls. 
%Another deceptive techniques, is to attempt to legitimate the project verifying the smart contract code on BSCscan and Etherscan and creating the "official" Telegram group of the token.
%We find evidence of this technique in the code of the smart contract token and then directly inspecting on Telegram. In particular we find the smart contract is available and verified for 55\% (147,069) of BSC and 67\% (14,722) of Ethereum tokens involved in the 1-day rug pull operations. Then, analyzing the source codes, we notice that 19,096 smart contract token in BSC and 1,334 in Ethereum reports link to the Telegram group of the token.
Another deceptive technique consists in attempting to legitimate the project by verifying the smart contract code on BSCscan and Etherscan.
The verification consists in uploading the source code so that the platform can compile it and verify that it matches the bytecode of the token stored in the blockchain. The verification provides users transparency and gives more guarantee that the token is not fraudulent. We find the smart contract is available and verified for 55\% (147,069) of BSC and 67\% (14,722) of Ethereum tokens involved in the 1-day rug pull operations.
Finally, another technique to legitimate the token consists in creating the "official" Telegram group of the token. We find evidence of this technique in the smart contract's code and then inspect the groups on Telegram. Indeed, analyzing the source codes, we notice that 19,096 token smart contracts in BSC and 1,334 in Ethereum report a link to the Telegram group of the token.
Although the organizers of 1-day rug pulls use deceptive techniques to dupe investors, we cannot consider these operations frauds because the phenomenon is still not regulated.
In any case, people lose money: Investors bought the token in 92.7\% of the Ethereum rug pulls and in 91.2\% of the BSC ones, and in all these cases the investment is lost. 
For this reason, we believe that these operations, even if not illegal, are exploitative of the DeFi ecosystem and should be contrasted to safeguard investors. 
Indeed, regulators are starting to take action to contrast them.
For example, New York State Senator Kevin Thomas proposes criminalizing rug pulls and other crypto frauds by introducing a new bill amendment request (Senate Bill S8839)~\cite{senatebill}. 
The idea of the bill is to introduce the crime of \textit{illegal rug pull} that occurs if the creator of the token sells more than 10\% of his tokens within five years of their last sale.

\section{Token names}
\begin{table}[h]
\small
    \centering
    \caption{%
       Token names most frequently used in 1-day rug pull operations. 
    }\label{tab:fakes}
    \begin{tabular}{@{\extracolsep{2pt}}l  r l r}
        \toprule
        \multicolumn{2}{c}{BNB Smart Chain}  & \multicolumn{2}{c}{Ethereum}\\
        \cmidrule{1-2} 
        \cmidrule{3-4} 
        Name & \# of tokens  & Name & \# of tokens \\
        \midrule
        Pornhub & 1,023  & Hyve.works  & 50 \\ 
        Galaxy & 588 & Deriswap & 32 \\ 
        Seedswap & 502 & Shibaswap & 28 \\ 
        Lionswap & 429 & Apple core finance  & 17 \\ 
        Eco.finance & 421  & X20.finance  & 16 \\ 
        Spacex & 419 & Yield farm rice & 15 \\ 
        Onlyfans & 419 & The sandbox  & 14 \\ 
    \end{tabular}
\end{table}
%%%%%%%%%%%%%%%%%%%%%%%%%%%%%%%%%%%%%%%%%%%%%%%%%%%%%%%%%%%%%%%%%%%%%%%%%%%%%%%%
\end{document}